\documentclass[a4paper,11pt,twocolumn]{article}    
\pdfoutput=1
\usepackage[T1]{fontenc}
\usepackage{graphicx} 
\usepackage{bbold}
\usepackage{mathtools}
\usepackage{amsmath}
\numberwithin{equation}{section} 
\usepackage{amsfonts}
\usepackage{amssymb}
\usepackage{amsbsy}
\usepackage{amsfonts}
\usepackage{physics}
\usepackage{bbold}
\usepackage{slashed} 
\usepackage{color}
\usepackage[table]{xcolor}
\usepackage{tablestyles}
\usepackage{tabularx}
\usepackage{hyperref}
\usepackage{bm}
\usepackage{cite}  
\usepackage{mathrsfs}
\usepackage{framed}
\usepackage[font=footnotesize,labelfont=bf]{caption}
\def\eq#1{{Eq.~(\ref{#1})}}

\def\Tr{\mbox{Tr}\,}

\newcommand{\di}{\mbox{d}}

\def\di{\mbox{d}}

\colorlet{grayline}{gray!70}
\definecolor{blueline}{rgb}{0,0.27,0.55}
\definecolor{DarkGray}{gray}{0.4}
\definecolor{Gray}{gray}{0.6}
\definecolor{oucrimsonred}{rgb}{0.6, 0.0, 0.0}
\definecolor{persianblue}{rgb}{0.11, 0.22, 0.73}
\definecolor{forestgreen}{rgb}{0.13,0.35,0.13}
 \hypersetup{colorlinks, citecolor=forestgreen, linkcolor=forestgreen, urlcolor=forestgreen}
\newcommand{\be}{\begin{equation}}
\newcommand{\ee}{\end{equation}}
\newcommand{\bea}{\begin{eqnarray}}
\newcommand{\eea}{\end{eqnarray}}

\newcommand{\CC}{\operatorname{C}}
\newcommand{\BB}{\operatorname{B}}
\newcommand{\hk}{{\bf \hat{k}}}
\newcommand{\hr}{{\bf \hat{r}}}
\newcommand{\hn}{{\bf \hat{n}}}
\newcommand{\hp}{{\bf \hat{p}}}

\newcommand*\xbar[1]{%
  \hbox{\;%
    \vbox{%
      \hrule height 0.5pt 
      \kern0.5ex
      \hbox{%
        \kern-0.25em
        \ensuremath{#1}%
        \kern-0.07em
      }%
    }%
  }%
} 
\newcommand{\com}[1]{}
\newcommand{\gsim}{\lower.7ex\hbox{$\;\stackrel{\textstyle>}{\sim}\;$}}
\newcommand{\lsim}{\lower.7ex\hbox{$\;\stackrel{\textstyle<}{\sim}\;$}} 

\newcommand{\bc}{\begin{center}}
\newcommand{\ec}{\end{center}}

\font\beeg=cmr17 scaled 1800
\newbox\ibox
\def\versal#1{\setbox\ibox=\hbox{{\beeg #1}~}%
	    \noindent\global\hangindent=\wd\ibox\global\hangafter-2%
	    \sc\smash{\llap {\lower 14pt \box\ibox}}}
\begin{document}
\onecolumn
\thispagestyle{empty}
\begin{center}
{ \Large \color{oucrimsonred} \textbf{ 
Dipole momenta and compositeness  \\[0.3cm] 
of the $\tau$ lepton at Belle II  
}}

\vspace*{1.5cm}
{\color{DarkGray}
  {\bf M. Fabbrichesi$^{a}$, and} 
 {\bf L. Marzola$^{{b}}$}
}\\

\vspace{0.5cm}
{\small 
{\it  \color{DarkGray} (a)
INFN, Sezione di Trieste, Via Valerio 2, I-34127 Trieste, Italy}
  \\[1mm]  
  {\it \color{DarkGray}
(b) Laboratory of High-Energy and Computational Physics, NICPB, R\"avala 10, \\ 10143 Tallinn, Estonia}
}
\ec

 \vskip0.5cm
\bc
{\color{DarkGray}
\rule{0.7\textwidth}{0.5pt}}
\ec
\vskip1cm
\bc
{\bf ABSTRACT}
\ec
The large number of $\tau$ leptons available at the Belle II experiment makes it possible to study their properties and the extent of their compositeness. Our strategy relies on three observables defined in terms of elements of the polarization density matrix of the produced $\tau$ pairs, which we obtain via  quantum tomography. The observables are able to explore values of the magnetic dipole moment down to $6.3 \times 10^{-4}$, which is two order of magnitude better than the current experimental limit, constrain the electromagnetic radius below $4.3 \times 10^{-3}$ fm and exclude an electric dipole moment larger than $1.7 \times 10^{-17}$ e cm. The quoted limits are at the $95\%$  confidence level and are obtained for a benchmark integrated luminosity of 1 ab$^{-1}$. 
\vspace*{5mm}

\noindent

  \vskip 3cm
\bc 
{\color{DarkGray} \vbox{$\bowtie$}}
\ec


\newpage
\section{Introduction\label{sec:intro}} 

{\versal The abundant production of $\tau$-lepton} pairs in $e^+ e^- \to \tau^+\tau^-$ at Belle II~\cite{Belle-II:2018jsg} provides a promising laboratory for a systematic analysis of their electromagnetic couplings. The study is  important because deviations of these parameters from their Standard Model (SM) values would provide insights into the nature of underlying new physics, implying for instance the presence of new particles or interactions. Given the current data, the $\tau$ lepton ---the heaviest among the leptons, as the top quark among the quarks---could be the likeliest to show a behavior departing from that described by the SM. In this paper we then investigate the electric dipole moment, the anomalous magnetic moment and the possible substructure of the $\tau$ lepton, represented by the size of its radius that we define below.

The most general electromagnetic Lorentz-invariant coupling between the photon and  the $\tau$ lepton can be written as
\be
-ie\,\bar \tau \,\Gamma^\mu(q^2) \,\tau \,A_\mu(q)
=
-ie \,\bar \tau \left[ \gamma^\mu F_1(q^2)  +  \frac{i \sigma^{\mu\nu}q_\nu}{2 m_\tau} F_2(q^2) 
+  \frac{\sigma^{\mu\nu} \gamma_5 q_\nu}{2 m_\tau} F_3(q^2)  \right]   \tau\, A_\mu(q) \, , \label{int}
\ee
which defines the  magnetic and electric dipole moments as
\be
a_\tau = F_2(0) \quad \mbox{and} \quad d_\tau = \frac{e}{2 m_\tau} F_3(0)\,,
\ee
as well as the  mean squared radius
\be
\label{rm}
\langle \vec r^{\;2} \rangle = -  6 \left .\frac{d G_E}{d \vec q^{\;2}} \right|_{q^2=0} \,,
\ee
with
\be
G_E(q^{2}) =  F_{1}(q^{2}) + \frac{q^{2}}{4 m_{\tau}^{2}} F_{2}(q^{2})\,. \label{G_E}
\ee
We take the form factors $F_{2,3}(q^2)$ at the leading order for $q^2\to 0$, while we retain the first order in $q^2$ for the form factor $F_1(q^2)$.

Whereas the magnetic moments of lighter leptons are measured to an accuracy that probes the respective radiative contributions expected within the SM, that of the heavier $\tau$ lepton is still only known to a lower precision: the current limits 
\be
-0.052  < a_\tau  <  0.013    \quad  \mbox{(95\%) CL} \; \text{\cite{DELPHI:2003nah}}\,,
\ee
are one order of magnitude above the theoretical value
\be
a_\tau^{\rm SM}=1.17721(5) \times 10^{-3}\,\text{\cite{Eidelman:2007sb}} \label{aSM}.
\ee

The electric dipole moment is constrained to be (with an integrated  luminosity of 833 fb$^{-1}$)
\be
-0.185  \times 10^{-16}     < d_\tau < 0.061 \times 10^{-16} \; \mbox{e cm} \quad \mbox{(95\%) CL}\; \text{\cite{Belle:2021ybo}}
\ee
assuming its value to be real. A non-vanishing value would signal the presence of CP violation.
 
No direct limits on the electromagnetic radius of the $\tau$ lepton exist, but there is a bound on the scale
\be
\Lambda_{\rm C.I.} > 7.9 \; \mbox{TeV} \; \text{\cite{ALEPH:2006jhv}}
\ee
for the related four-fermion contact interaction 
\be
+\frac{2\pi}{\Lambda^{2}_{\rm C.I.}} \, (\bar e_{L} \gamma^{\mu}e_{L}) \, (\bar \tau_{L}\gamma_{\mu} \tau_{L}) \,,
\ee
constrained at the energy of about 200 GeV.

The form factors introduced in \eq{int} translate, in the language of  effective field theory, into  the $SU(2)\times U(1)$ invariant effective operators for the $\tau$ lepton. The leading contributions come from the following three higher-dimensional operators:
\be\label{O}
\hat O_1 = e \frac{c_1}{m_\tau^2} \bar \tau \gamma^\mu  \tau D^\nu F_{\mu\nu} \, ,\quad  \hat O_2= e \frac{c_2\,\upsilon}{2 m_\tau^2 } \bar \tau \sigma^{\mu\nu} \tau F _{\mu\nu}\quad \mbox{and} \quad\hat O_3= e \frac{c_3\,\upsilon}{2 m_\tau^2 } \bar \tau \sigma^{\mu\nu} \gamma_{5}\tau F _{\mu\nu}\, ,
\ee
where $D^\nu$ is the covariant derivative, $F_{\mu\nu}$ the electromagnetic field strength tensor and $\upsilon=174$ GeV is the Higgs field vacuum expectation value. In \eq{O}  the dimensionless Wilson coefficients $c_i$ are taken to be real. Left- and right-handed chiral fields enter symmetrically. The operator $\hat O_1$ gives the leading $q^2$ dependence of the form factor $F_1$, while $\hat O_{2,3}$ give the $q^2$-independent term of the form factors $F_{2,3}$:
\be 
\label{f12}
F_1(q^2)=1+c_1 \frac{q^2}{m_\tau^2} +\ldots\qquad\mbox{and}\qquad F_{2,3}(0)=2\, c_{2,3} \frac{\upsilon}{m_\tau} \,.
\ee

The operators $\hat O_{2,3}$ are written for the sake of convenience with an extra factor $\upsilon/m_\tau$, sourced after the electroweak symmetry breaking by the dimension-six $SU(2)_L\times U(1)_Y$ gauge invariant operators involving the Higgs field. 
Operators of higher dimensions can in general contribute---they give further terms in the expansion of the form factors---but their effect is suppressed. 

Compared to the effective field theory approach, the form factors we use have the advantage of directly answering  a simple question: is the $\tau$ lepton  a point like particle or does it show a composite structure? The answer comes without relaying on any assumption or UV model to select the relevant operators out of the many present in the  effective theory.

Our strategy to constrain the electromagnetic couplings in \eq{int} exploits the polarization density matrix, which can be experimentally reconstructed through a procedure dubbed \textit{quantum tomography} and gives a bird's-eye view of the possible observables available for a given process. The method has been previously used to constrain physics beyond the SM affecting the top-quark~\cite{Aoude:2022imd,Fabbrichesi:2022ovb} and $\tau$ pair~\cite{Fabbrichesi:2022ovb} production at the LHC, or yielding Higgs anomalous couplings to $\tau$ leptons~\cite{Altakach:2022ywa} and gauge bosons~\cite{Fabbrichesi:2023jep,Bernal:2023ruk,Aoude:2023hxv}.
For the present case, we use the entries of the polarization density matrix to define three observables that provide the means to best constrain the parameters in \eq{int}: one observable measures the entanglement~\cite{Horodecki:2009zz} in the spin states of the produced $\tau$ pairs, another is related to triple products involving one momentum and the spin vectors of the $\tau$ leptons---and it is specific to the CP-violating electric dipole moment, and the third is the total cross section.  

We compare our results to the current experimental limits~\cite{Workman:2022ynf} and to recent phenomenological estimates of the electric and magnetic dipole moments in~\cite{Chen:2018cxt} and of the electric dipole moments in~\cite{Bernreuther:2021elu}.

\subsection{Events}
 
We propose to probe the electromagnetic couplings of the $\tau$ lepton by using the process $e^+ e^- \to \tau^+\tau^-$  at the Belle-II experiment, located at the SuperKEKB collider. The SuperKEKB collider delivers $e^+ e^- $ collisions at a center of mass energy of $\sqrt{s} = 10.579$ GeV. The Belle collaboration has published analyses of $e^+ e^- \to \tau^+\tau^-$  production with data corresponding to an integrated luminosity of up to $921$~fb$^{-1}$~\cite{Belle:2020lfn}, equivalent to $841$ million $ \tau^+\tau^-$ events, and other 175 million events have now been added with the Belle II dataset~\cite{Belle-II:2023izd}. 
The aim of the SuperKEKB project is to collect $50$~ab$^{-1}$ of data~\cite{Akai:2018mbz,Belle-II:2018jsg}, corresponding to a dataset of about $50$ billion $e^+ e^- \to \tau^+\tau^-$ events. Of these, we can use those  involving the $\tau^- \to \pi^-\nu_\tau$, $\tau^- \to \pi^-\pi^0\nu_\tau$ and $\tau^- \to \pi^-\pi^+ \pi^- \nu_\tau$ decays to reconstruct the polarization density matrix. The combination of these channels covers about $21\%$ of $\tau$ pair decays.

We expect the dominant background to arise from mis-reconstructions of the $\tau$ decay channel, affecting about $15\%$ of the $e^+ e^- \to \tau^+\tau^-$ events at Belle II~\cite{Belle-II:2023izd}. Backgrounds arising from the process $e^+ e^- \to \textrm{q}\bar{\textrm{q}}$ and from other sources are negligible in comparison. Given their modest impact, we neglect the effect of backgrounds in our study. It will have to be taken into account when the analysis is performed on the actual data.

The detailed Monte Carlo simulation of the process $e^+ e^- \to \tau^+\tau^-$ in~\cite{Ehataht:2023zzt} shows that a full quantum tomography is achievable and provides an estimate of the uncertainties in the density matrix that we are going to use in our analysis.

\section{Methods} 
{\versal Quantum tomography aims} to fully determine the density matrix $\rho$ of a quantum state. The $\tau$ leptons---whose spins are represented with two-level quantum states, that is, \textit{qubits}---act as their own polarimeters and the full polarization density matrix can be reconstructed, within the inherent uncertainties of the procedure, from the angular distribution of the $\tau$ decay products.

Polarizations are more difficult to measure than momenta and so the reconstruction of the polarization density matrix from the data is challenging. The main aim of this work is to show to what extent the advantages of using the proposed method make the extra work in the experimental analysis eventually worthwhile. 

The density matrix describing the polarization state of a quantum system composed by two fermions can be written as
	\begin{equation}
	\label{eq:rho_deco}
		\rho = \frac{1}{4} \qty[
		\mathbb{1}\otimes\mathbb{1} 
		+ 
		\sum_i \BB_i^+ \, \qty(\sigma_i \otimes \mathbb{1}) 
		+ 
		\sum_j \BB_j^- \, \qty(\mathbb{1} \otimes \sigma_j )
		+
		\sum_{i,j} \CC_{ij} \, \qty(\sigma_i \otimes \sigma_j)
		],
	\end{equation}
where $i,j\in\{n,r,k\}$ and $\sigma_i$ are the Pauli matrices. The decomposition refers to a right-handed orthonormal basis, $\{\hn, \hr, \hk\}$ and the quantization axis for the polarization is taken along $\hk$, so that $\sigma_k\equiv\sigma_3$. In the fermion-pair center of mass frame we have
	\begin{equation}
		\hn = \frac{1}{\sin(\theta)}\qty(\hp \times\hk), \quad \hr = \frac{1}{\sin(\theta)}\qty(\hp-\cos(\theta)\hk)\,,
	\end{equation}
where $\hk$ is the direction of the $\tau^+$ momentum and $\theta$ is the scattering angle satisfying $\hp\cdot\hk = \cos\theta$, with $\hp$ being the direction of the incoming $e^+$. 

The coefficients $\BB^\pm_i$ in \eq{eq:rho_deco} give the polarizations of the individual fermions, whereas the matrix $\CC_{ij}$ contains the spin correlations. By using the properties $\Tr(\sigma_i\sigma_j)=2\delta_{ij}$ and $\Tr(\sigma_i)=0$ we have:
	\be
		\BB_i^+ = \frac{\Tr[\rho\qty(\sigma_i \otimes \mathbb{1}) ]}{4}\,,\quad
		\BB_i^- = \frac{\Tr[\rho\qty( \mathbb{1} \otimes \sigma_i) ]}{4}\,,\quad
		\CC_{ij} = \frac{\Tr[\rho\qty(\sigma_i \otimes \sigma_j)]}{4}\,,
	\ee
and, for the process at hand, it holds that $\BB^\pm_i=0$.

The non-vanishing coefficients $\CC_{ij}$ can be reconstructed in the actual experiments by tracking the angular distribution of suitable $\tau$-pair decay products. In particular, for events where each $\tau$ lepton decays to a single pion and a neutrino, we have
\begin{equation}
  \frac{1}{\sigma} \, \frac{d\sigma}{d\cos\theta^+_{i} \, d\cos\theta^-_{j}} = \frac{1}{4} \, \left( 1 + \CC_{ij} \, \cos\theta^+_{i} \, \cos\theta^-_{j} \right) \, ,
  \label{eq:xsec2d}
\end{equation}
where $\cos\theta^\pm_{i}$ is the projections of the $\pi^\pm$ momentum direction on the $\{\hn, \hr, \hk\}$ basis, as computed in the rest frame of the decaying $\tau^\pm$. Crucial to the whole procedure is the reconstruction of the neutrino kinematics. More details pertaining to the experimental determination of the $\CC_{ij}$ coefficients for different decay channels are presented, for instance, in Ref.~\cite{Ehataht:2023zzt}, which we closely follow.     

Quantum tomography gives the coefficients $\BB_i$ and $\CC_{ij}$ for the density matrix and there is a number of observables that can be constructed with them. We consider the three that provide the most stringent limits on the anomalous couplings: 
\begin{itemize}
\item
The concurrence  $\mathscr{C}[\rho]$~\cite{Horodecki:2009zz},  which  can be computed by means of the particularly simple formula 
\be
\mathscr{C}[\rho] = \frac{1}{2} \max \Big[0, \, |\CC_{rr}+\CC_{kk}| -(1+\CC_{nn})\,,\,  \sqrt{(\CC_{rr}-\CC_{kk})^2 + 4 \CC_{rn}^2} - |1 - \CC_{nn}| \Big] \, ,\label{Crho}
\ee
because the $B_i^\pm=0$ and all off-diagonal elements but $\CC_{rn}$ vanish when the scattering angle is integrated over. We use as observable the concurrence \eq{Crho} computed from the $\CC$ matrix obtained by averaging over the angular distribution of the $\tau$ leptons.

\item The second observable is just the total cross section:
\be
\sigma = \frac{1}{64 \pi^{2}\, s}\int \di \Omega \frac{|{\cal M} |^{2}}{4} \sqrt{1 - \frac{4 m_\tau^2}{s}}\,,
\ee
where we neglect the electron mass and $\sqrt{s}=10.579$ GeV. The spin-summed squared amplitude, $|{\cal M}|^{2}$, is given by
\begin{multline}
  |{\cal M}|^{2} = 
   e^4\Bigg\{
    \frac{s}{m_\tau^2}\left[\sin^2\theta \left(16 \, c_1^2+\tilde{d}_\tau^2+a_\tau^2\right)+4\, c_1 \left(\cos 2\theta + 4\, a_\tau+3\right)\right]
     +2\, \frac{s^2}{m_\tau^4}\, c_1^2 \Big(\cos 2 \theta +3\Big)
     \\
     +4 \left(8 \,c_1 \,\sin ^2\theta
     +\left(\tilde{d}_\tau^2+a_\tau^2+1\right) \cos^2\theta -\tilde{d}_\tau^2+a_\tau^2+4 \,a_\tau+1\right)
     +16\, \frac{m_\tau^2}{s} \sin ^2\theta \Bigg\}\,.   
\end{multline}
\item
The third observable, $\mathscr{C}_{odd}$, singles out the antisymmetric parts of the density matrix. It is defined as
\be
\mathscr{C}_{odd}=\frac{1}{2}\, \sum_{\substack{i, j\\ i< j}} \Big| \CC_{ij} -\CC_{ji} \Big| \, , \label{CPodd} 
\ee
and contains only off-diagonal terms that change sign under transposition. The observable encodes kinematical variables that can be written as the triple products of momenta and spin vectors, for instance:
\be
\vec k \cdot \Big(\vec{s}_{\hat n} \times \vec{s}_{\hat r} \Big) \, , \label{Todd}
\ee
with $\vec k$ the momentum of one of the particles and $\vec{s}_{\hat n}$ and  $\vec{s}_{\hat r}$ the projections of the spin vector along two directions orthogonal to the momentum. In the present case there is only one non-vanishing term coming from the elements $\CC_{rn}$ and $\CC_{nr}$, which are equal in magnitude and with the opposite sign. We build the observable $\mathscr{C}_{odd}$ from the $\CC$ matrix obtained by averaging over the angular distribution of the $\tau$ leptons. 
\end{itemize}
The integration over the angular distribution used for the operators $\mathscr{C}[\rho]$ and $\mathscr{C}_{odd}$ is performed over the range defined by $|\cos \theta| < 0.4$ to optimize the quantum tomography procedure.

\subsection{Uncertainties}

To set limits on the parameters in \eq{int} we need to know the uncertainties of the operators we utilize to characterize the process $e^+e^-\to \tau^+\tau^-$ at Belle II. These have been estimated for a benchmark luminosity of 1 ab$^{-1}$ as follows. For the concurrence $\mathscr{C}[\rho]$ and the CP odd operator $\mathscr{C}_{odd}$ we rescale the corresponding uncertainties of $1.4 \times 10^{-3}$ and $4.0 \times 10^{-3}$ given in Ref.~\cite{Ehataht:2023zzt} (after averaging on the angular distribution of $\tau$ leptons) for a luminosity of 220 fb$^{-1}$ and for a center of mass energy $\sqrt{s}=10.579$ GeV. For the total cross section, instead, we rescale the relative uncertainty of 0.3\% on the integrated luminosity quoted in Ref.~\cite{Banerjee:2007is} for 833 fb$^{-1}$. All these uncertainties are at the $1\sigma$ confidence level and contain (to different extents) also systematic errors.

\section{Results} 
\begin{figure}[h!]
\begin{center}
\includegraphics[width=2.7in]{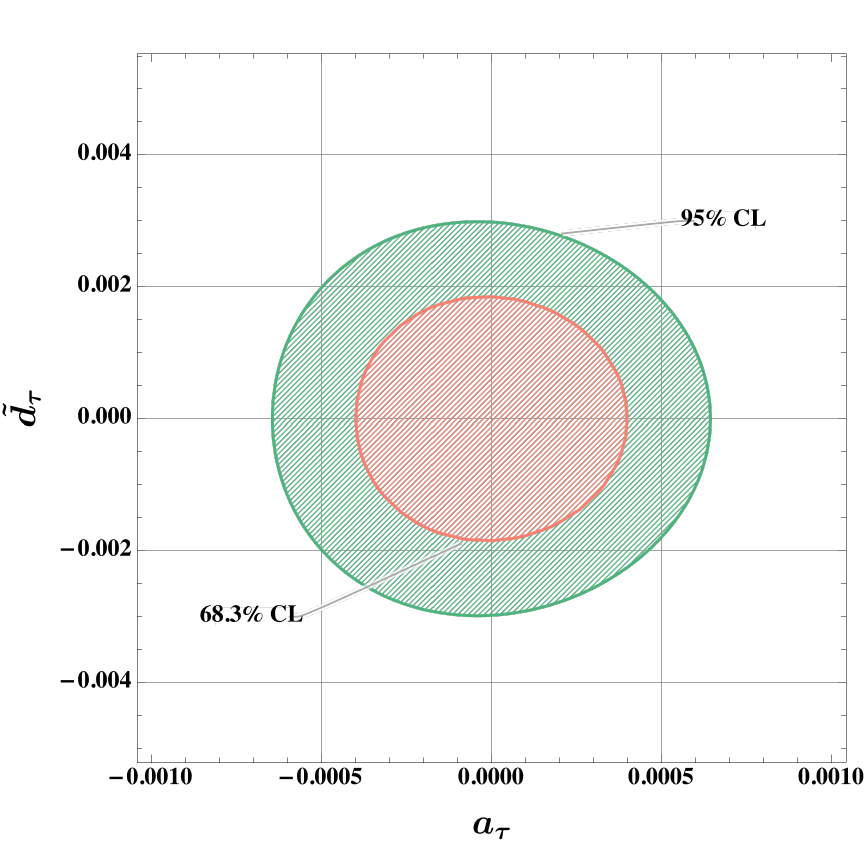}
\hspace{.5cm}
\includegraphics[width=2.7in]{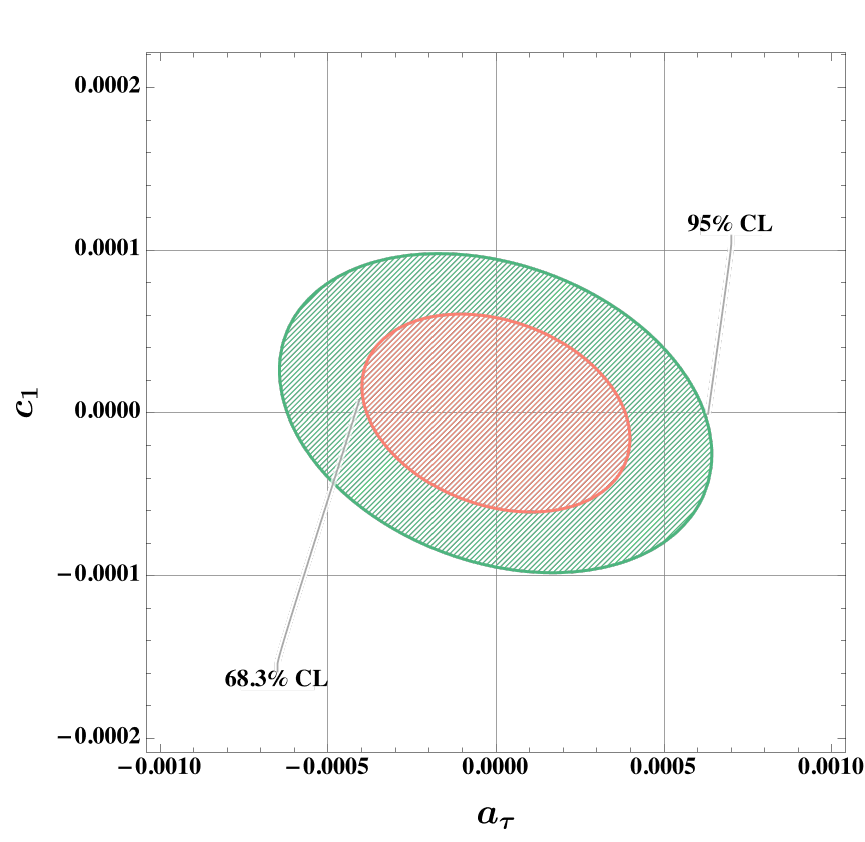}
\caption{\footnotesize 
\label{fig:limits1} Bounds on the electromagnetic couplings $a_{\tau}$, $\tilde d_{\tau}= F_3(0)$ and $c_{1}$ obtained by means of \eq{chi2}  varying two parameters at a time.
}
\end{center}
\end{figure}

{\versal The operators introduced} in the previous section, generically denoted here as  $\mathscr{O}_i(a_\tau,d_\tau,c_1)$, depend on the electromagnetic couplings $a_\tau$, $d_\tau$ and $c_1$, and $\mathscr{O}_i(0,0,0)$ are the  values of the operators for the SM. To constrain the couplings, we introduce a $\chi^2$ test set for a (68.3) 95\% joint CL
\be
\sum_{i} \left[ \frac{\mathscr{O}_i(a_\tau,d_\tau,c_1)-\mathscr{O}_i(0,0,0)}{\sigma_{i}}\right]^2 \leq (2.30)\; 5.99 \,,  \label{chi2}
\ee
in which we set the uncertainties $\sigma_i$ for the operators $\mathscr{O}_i$ at the values discussed in the previous section.

The operators $\mathscr{O}_{i}$ in \eq{chi2} that we utilize to constrain $a_{\tau}$ and $d_{\tau}$ are  $\mathscr{C}[\rho]$ and $\mathscr{C}_{odd}$. We employ instead $\mathscr{C}[\rho]$ and the cross section $\sigma$ to constrain $a_{\tau}$ and $c_{1}$.

The bounds on each coupling can be extracted from Fig.~\ref{fig:limits1} via marginalization by assuming the parameters to be independent. The values obtained from the 95\% joint confidence interval are reported in Tab.~\ref{tab:couplings}, where they are compared to the current experiment bounds.

\begin{table}[t]
  \tablestyle[sansboldbw]
  \begin{tabular}{*{2}{p{0.35\textwidth}}}
  \theadstart
      \thead PDG (2022) &\thead  This work \\
  \tbody
 $-1.9 \times 10^{-17} \leq d_{\tau} \leq 6.1 \times 10^{-18} $ e cm& $  |d_{\tau}| \leq 1.7 \times 10^{-17} $ e cm    \\
 $ -5.2 \times 10^{-2} \leq a_{\tau} \leq 1.3 \times 10^{-2}$ &  $ | a_{\tau} |\leq 6.3 \times 10^{-4} $    \\
 $\qquad \quad \Lambda_{\rm C.I.}\geq 7.9\; \text{TeV}$ & $ |c_{1}| \leq 9.5 \times 10^{-5}\, , \quad  \Lambda_{\rm C.I.}\geq 2.6\; \text{TeV}$    \\
  \hline%
  \tend
  \end{tabular}
  \caption{\footnotesize \label{tab:couplings} \textrm{Bounds obtained on the single parameter after marginalization of the 95\% joint confidence intervals for the electromagnetic couplings shown in Fig.~\ref{fig:limits1}, neglecting correlations. The values refer to a luminosity of 1 ab$^{-1}$ at Belle II; limits for higher luminosities can be (approximately) obtained by rescaling these values by the square root of the ratio of the relative luminosities. The current experimental limits are reported in the first column.}}
  \end{table}

\begin{figure}[h!]
\begin{center}
\includegraphics[width=3.5in]{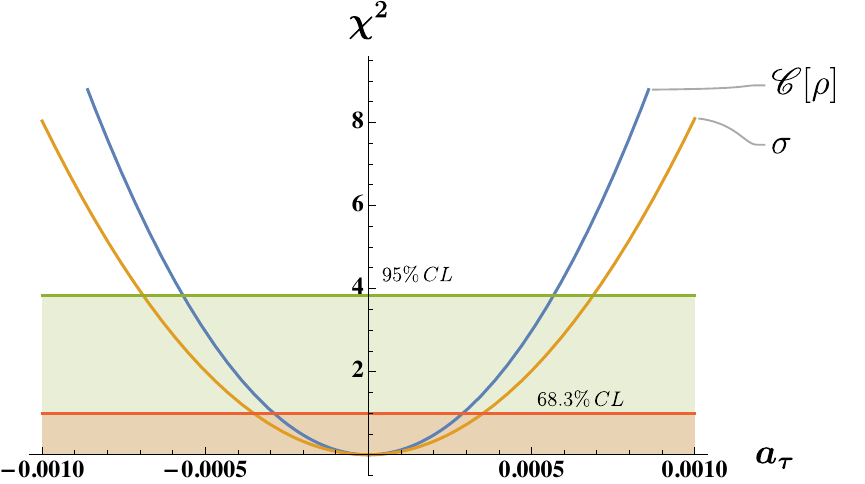}
\caption{\footnotesize 
\label{fig:limits2} Comparison between the power of entanglement (blue line) and cross section (orange line) to constrain the electromagnetic couplings. For the plot we use a common relative uncertainty of $0.1\%$. The green and red areas denote the 95\% and 68.3\% one-parameter confidence level, respectively.
}
\end{center}
\end{figure}
The concurrence provides a limit 20\% stronger than that given by the total cross section in determining the anomalous magnetic moment, as shown in Fig.~\ref{fig:limits2}. Whether this improvement in the constraint justifies the extra work required by quantum tomography depends on the details of the experimental setup. Certainly, the method is not beneficial if the error in the cross section measurement is much smaller than the corresponding uncertainty in the determination of the $\tau$ lepton polarizations.

The radius of the $\tau$ lepton that can be explored is estimated from our limit on $c_1$ and \eq{G_E} as
\be
\langle \vec r^{\;2} \rangle = \frac{6}{m_\tau^2} \, \left(c_1 + \frac{a_\tau}{4}\right)\,,
\ee
which gives the root mean squared radius value of
\be
\sqrt{\langle \vec r^{\;2} \rangle} <  5.1 \times 10^{-3} \; \text{fm.}
\ee
By comparison, the electron, which is believed to be a point-like particle, has a limit on its root mean squared radius of about $10^{-5}$ fm~\cite{Bourilkov:2001pe} that does not originate in radiative corrections. 

\section{Outlook}

{\versal We have outlined a strategy} to constrain the electric and magnetic dipole moments of the $\tau$ lepton, as well as its electromagnetic radius, via quantum tomography.

In order to put our results into perspective, we compute the single-parameter 68.3\% confidence intervals for $a_\tau$ and $d_\tau$ from \eq{chi2} and compare the resulting limits to those in the Literature. With a luminosity of 1~ab$^{-1}$, Ref.~\cite{Chen:2018cxt} finds 
\be
|d_\tau -d_\tau^{SM}| < 1.44 \times 10^{-18} \; \text{e cm} \quad \text{and} \quad |a_\tau -a_\tau^{SM}|< 1.24 \times 10^{-4}
\ee
at 68.3\% CL. Our method instead yields
\be
|d_\tau | < 6.7 \times 10^{-18} \; \text{e cm} \quad \text{and} \quad a_\tau < 2.6 \times 10^{-4}\, .
\ee
The SM prediction for $d_\tau^{SM}$ is still negligible at this precision level and so the first limit can be compared directly to ours. Differently, the SM prediction for $a_\tau^{SM}$ is given in \eq{aSM} and therefore the limit in~\cite{Chen:2018cxt} is about one order of magnitude weaker than our result. 

Ref.~\cite{Bernreuther:2021elu} quotes the following 68.3\% confidence interval for the electric dipole moment
\be
|d_\tau | < 6.8 \times 10^{-20} \; \text{e cm} \label{Bern}\,,
\ee
given for a luminosity of 50 ab$^{-1}$. Rescaling the result we obtained to account for the latter, we find 
\be
|d_\tau | < 9.4 \times 10^{-19} \; \text{e cm}  \, .
\ee

It is not surprising that both the limits on the electric dipole moment reported in Ref~\cite{Bernreuther:2021elu} and~\cite{Chen:2018cxt} are comparable to ours: these results are all obtained by using an operator involving the triple product of two spin vectors and one momentum, equivalent to the CP odd operator utilized in our analysis. The limit in \eq{Bern} is stronger because of the smaller uncertainty used, which is obtained by combining in quadrature the yields of different decay channels. Our result, in fact, is comparable to those obtained in Ref.~\cite{Bernreuther:2021elu} for any single channel.
  
As for the bound obtained for the compositeness scale, see Tab.~\ref{tab:couplings}, one has to bear in mind that the current experimental limit is obtained considering energies of about 200 GeV, whereas ours is given for energies one order of magnitude smaller, of about 10 GeV. Depending on the origin of the relevant four-fermion interaction, the scaling of the related operator is at least linear in the energy, implying that our result must be compared with the approximated rescaled value $\Lambda_{\rm C.I.} \gtrsim 800 \; \text{GeV}$ when assessing the power of the method.  

\section*{Acknowledgements}
{\small
We are happy to thank E.~Gabrielli, C.~Veelken and L.~Zani for discussions. L.M. is supported by the CoE grant TK202 ``Universum''.}

\twocolumn  
\small
\bibliographystyle{JHEP}   
\bibliography{taus.bib}

\end{document}